\documentclass[pra,aps,twocolumn,showpacs]{revtex4}
\usepackage{graphicx,epsf,bm,bbm,color,amsmath,ulem,amssymb,psfrag}

	\newcommand{\beq}{\begin{equation}}
	\newcommand{\be}{\begin{equation}}
	\newcommand{\beqn}{\begin{eqnarray}}
	\newcommand{\eeq}{\end{equation}}
	\newcommand{\ee}{\end{equation}}
	\newcommand{\eeqn}{\end{eqnarray}}

\newcommand{\bem}{\begin{pmatrix}}
\newcommand{\eem}{\end{pmatrix}}

\begin{document}

	\today
	
		\title{Inter-band tunneling near the merging transition of Dirac cones}
        \author{Jean-No\"el Fuchs}
       \email{fuchs@lptmc.jussieu.fr}
        \affiliation{Laboratoire de Physique des Solides, CNRS UMR 8502, Univ. Paris-Sud, F-91405 Orsay}
        \affiliation{Laboratoire de Physique Th\'eorique de la Mati\`ere Condens\'ee, CNRS UMR 7600, Universit\'e Pierre et Marie Curie, 4 place Jussieu, F-75252 Paris}
        \author{Lih-King Lim}
         \affiliation{Laboratoire de Physique des Solides, CNRS UMR 8502, Univ. Paris-Sud, F-91405 Orsay}
        \affiliation{Laboratoire Charles Fabry, Institut d'Optique, CNRS UMR 8501, Universit\'e Paris-Sud, 2 av. Fresnel, F-91127 Palaiseau}
        \author{Gilles Montambaux}
        \affiliation{Laboratoire de Physique des Solides, CNRS UMR 8502, Univ. Paris-Sud, F-91405 Orsay}

		\begin{abstract}
Motivated by a recent experiment in a tunable graphene analog [L. Tarruell {\it et al.}, Nature \textbf{483}, 302 (2012)], we consider a generalization of the Landau-Zener problem to the case of a quadratic crossing between two bands in the vicinity of the merging transition of Dirac cones. The latter is described by the so-called universal hamiltonian. In this framework, the inter-band tunneling problem depends on two dimensionless parameters: one measures the proximity to the merging transition and the other the adiabaticity of the motion. Under the influence of a constant force, the probability for a particle to tunnel from the lower to the upper band is computed numerically in the whole range of these two parameters and analytically in different limits using (i) the St\"uckelberg theory for two successive linear band crossings, (ii) diabatic perturbation theory, (iii) adiabatic perturbation theory and (iv) a modified St\"uckelberg formula. We obtain a complete phase diagram and explain the presence of probability oscillations in terms of interferences between two poles in the complex time plane. We also compare our results to the above mentioned experiment.
		\end{abstract}
	\maketitle
	%\showpacs

\section{Introduction}
The recent surge of interest in bandstructure of unusual crystals spurred from various developments in studying condensed matter systems. To name two examples, relevant to the present work, are the successful isolation of single layer graphene \cite{Neto} and the fabrication of 3D topological insulators \cite{Hasan,Qi}. In contrast to ordinary two-dimensional (2D) crystals, the low-energy hamiltonian involves at least two coupled bands. It gives rise to band crossings which, depending on the material parameters, can be gapped or not. The extra degree of freedom in the internal space (generically called the pseudospin space) offers opportunity for the investigation of new system properties.

Most recently, hamiltonian engineering with artificial crystals   -- which is a main theme of the active field of quantum simulation, see e.g. \cite{BDN} -- provides a complementary route to realize coupled-band systems \cite{Mortessagne,Tarruell,Gomes}. While the low-energy hamiltonian mimics closely that of its solid-state counterpart, it is no longer limited to parameters of the actual material. For example, in the cold atom experiment performed at ETH Z\"{u}rich \cite{Tarruell}, a Dirac cone pair in the bandstructure is brought to merge as a function of laser parameters, thus realizing a Lifshitz transition which has never been reached in graphene \cite{Hasegawa,Dietl,Wunsch,MontambauxUH}. The merging is a topological transition in which two Dirac cones of opposite Berry phase approach and annihilate before a gap opens.

In the ETH Z\"urich experiment, Bloch oscillations of non-interacting fermionic atoms in a honeycomb-like optical lattice are executed to study the merging transition of Dirac cones. As the atom traces out a closed trajectory in the momentum space, it may tunnel to the second band when it comes close to a linear avoided band crossing (i.e. a Dirac cone), a process known as Landau-Zener tunneling. By measuring the transfer probability after performing a Bloch cycle, information about the bandstructure can be extracted with momentum resolution \cite{Weitz}.

\begin{figure}[ht]
\begin{center}
\includegraphics[width=5cm]{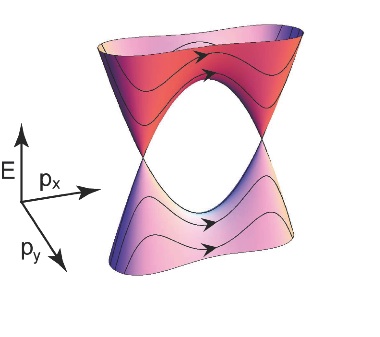}
\end{center}
 \caption{(Color online) Energy spectrum in the gapless phase ($\Delta_*<0$): energy $E=\pm \sqrt{(p_x^2/(2m^*)+\Delta_*)^2+c_y^2p_y^2}$ as a function of momentum $p_x\sim t$ and $p_y \sim k$. The distance between the two Dirac cones is controlled by the merging gap $\Delta_*\propto d$. The perpendicular gap $c_y p_y \propto k$ controls how far the particle is from hitting the Dirac cones, which are located at $(t= \pm \sqrt{|d|},k=0)$, directly at their tip. The black lines are lines of constant $k$.}
\label{fig:zenery}
\end{figure}

In Ref. \cite{Lim}, we presented a tight-binding model that reproduces the optical band structure in the parameter space of the experiment. Using a low-energy description of the tight-binding model known as the universal hamiltonian \cite{MontambauxUH}, we quantitatively reproduce the experimental results of Ref. \cite{Tarruell}. In the framework of the universal hamiltonian, the inter-band tunneling problem depends only on two relevant parameters: the merging gap $d$ -- which controls the proximity to the transition -- and the momentum perpendicular to the direction of motion $k$ -- which controls the adiabaticity of the motion, or in other words how far in momentum space the atom is from hitting exactly the tip of the Dirac cone. In particular, in Ref. \cite{Lim} we explain the situation where the two Dirac cones are hit in succession during a single Bloch oscillation (see Fig. \ref{fig:zenery}) by using a simple approximation, known as the St\"uckelberg theory \cite{Stuckelberg,revueStuck}, in which tunneling events are assumed to be independent. The validity of this approach is restricted to the gapless phase ($d<0$) and not too close to the merging transition ($d\ll -1,-k$), i.e. the two Dirac cones are well separated.

The present paper is an extension of our letter \cite{Lim} and focuses on the tunneling problem where the atom encounters two Dirac cones in succession. Here, we go beyond the independent cone approximation and present a complete picture of the inter-band transition probability as a function of the two parameters $d$ and $k$. In particular, we now access the whole phase diagram, including the gapped phase ($d>0$) and the transition point ($d=0$). Our paper is organized as follows. In section II, we formulate the inter-band tunneling problem for the universal hamiltonian. In section III, we recall the approximate solution used in \cite{Lim} based on the St\"uckelberg theory. We then present three other analytical approaches: diabatic perturbation theory in section IV, adiabatic perturbation theory in section V and a modified St\"uckelberg formula in section VI. In section VII, we present numerical solutions in the whole parameter space and compare the results of the different approaches. Finally in section VIII, we compare our results to the ETH Z\"urich experiment before concluding in section IX.

%------------------------------
%------------------------------

\section{Inter-band tunneling in the universal hamiltonian}
In the Landau-Zener (LZ) problem \cite{Landau,Zener,Wittig}, an avoided linear crossing between two bands is considered and the probability for a particle (which will be called an electron in the following) to tunnel from the lower to the upper band under a constant applied force is calculated. Landau solved the problem approximately using perturbation theory and the semiclassical approximation \cite{Landau}, while Zener was able to find the exact solution \cite{Zener}. For a concise modern presentation, see Ref. [\onlinecite{Wittig}] . Here we consider such a tunneling problem for the case of a quadratic band crossing. The latter occurs close to the merging transition of Dirac points \cite{Hasegawa,Dietl,Wunsch,MontambauxUH} and was recently observed in a cold atom realization of a graphene analog \cite{Tarruell}. In the gapless phase, the quadratic band crossing can be approximated as two successive linear crossings (or Dirac cones), which is at the heart of the St\"uckelberg approach (see below).

We start from the universal hamiltonian describing the vicinity of the merging transition \cite{MontambauxUH}:
\beq
H_u=\left[\frac{p_x^2}{2m_*} + \Delta_*\right] \sigma_x + c_y p_y \sigma_y \label{Hu}
\eeq
It depends on three real parameters: an effective mass $m_*> 0$ giving the spectrum curvature in the $x$ direction, an effective velocity $c_y>0$ for the $y$ direction and a merging gap $\Delta_*$, which is a real number controlling the distance to the transition \cite{xy}. The state space is that of an electron moving in a two-dimensional plane and carrying a pseudo-spin 1/2 described by the Pauli matrices $\sigma_x,\sigma_y,\sigma_z$. The corresponding spectrum is $E=\pm \sqrt{(p_x^2/2m_*+\Delta_*)^2+c_y^2p_y^2}$ and is plotted in Fig. \ref{fig:zenery} when $\Delta_*<0$. If the merging gap is negative, the spectrum is gapless and contains two Dirac cones at $(p_x=\pm \sqrt{2m_* |\Delta_*|},p_y=0)$. If it is zero (the merging point), the two Dirac cones are on top of each other and the spectrum is linear in one direction and quadratic in the perpendicular direction $E=\pm \sqrt{(p_x^2/2m_*)^2+c_y^2p_y^2}$ \cite{Dietl}. If it is positive, there are no band touching points anymore but a true gap $2\Delta_*$ between the two bands.

We add a constant electric field $\mathcal{E}$ in the $x$ direction such that during its motion an electron encounters the two Dirac cones in succession \cite{nosinglecone}, see Fig. \ref{fig:zenery}. The gauge is such that the vector potential $A_x=-\mathcal{E}t$ and $A_y=0$. Therefore
\beq
H_u(t)=\left[\frac{(p_x-Ft)^2}{2m_*} + \Delta_*\right] \sigma_x + c_y p_y \sigma_y \label{Hu2}
\eeq
The force $F=e\mathcal{E}$ is taken to be positive and $-e<0$ is the electron charge. The hamiltonian commutes with $p_x$ and $p_y$ and therefore the non-trivial dynamics only occurs in the internal space of the pseudo-spin $1/2$ and $p_x$ and $p_y$ can be taken as c-numbers (conserved quantities). Shifting the origin of time $Ft-p_x\to Ft$, it is now possible to get rid of $p_x$. This hamiltonian defines a characteristic energy scale   $E_{char}=(\hbar F)^{2/3}/(2m_*)^{1/3}$, and therefore a  timescale $t_{char}=\hbar/E_{char}$ and a length scale $L_{char}=E_{char}/F$. Energies, times and lengths are therefore given in units of these characteristic scales. We then define the dimensionless quantities $d\equiv \Delta_*/E_{char}$ and $k\equiv c_y p_y/E_{char}$ and the dimensionless hamiltonian  $H_u(t)=[t^2+d]\sigma_x+ k \sigma_z$. 

\begin{figure}[ht]
\begin{center}
\includegraphics[width=6cm]{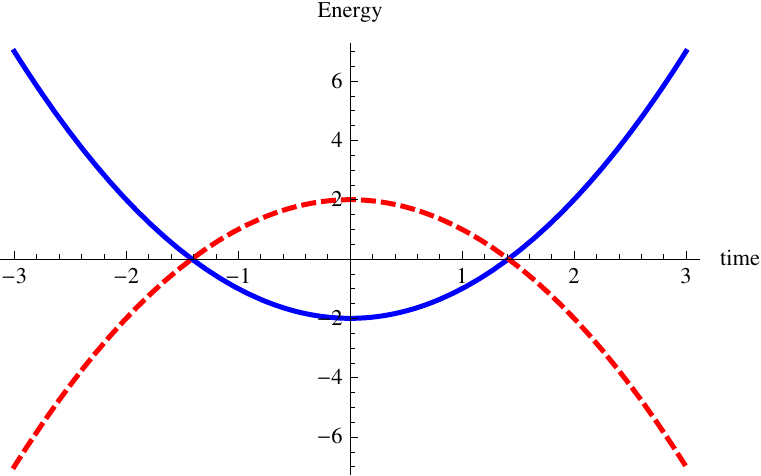}
\includegraphics[width=6cm]{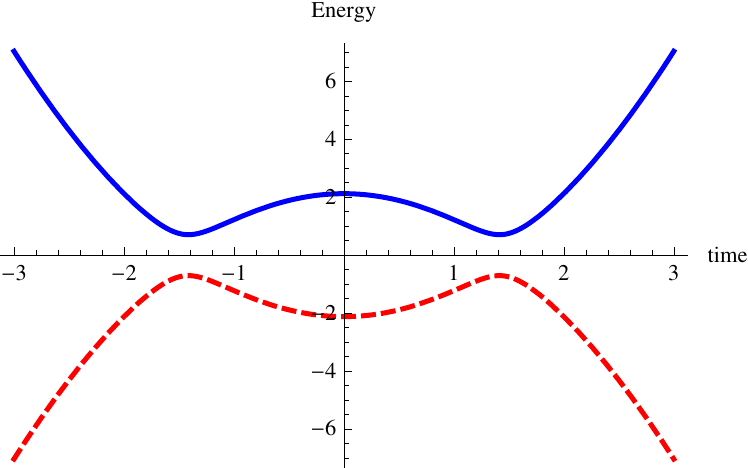}
\end{center}
 \caption{(Color online) Energy $E$ as a function of time $t$ when $d=-2<0$. (a) Diabats $E_{1,2}=\pm [t^2 +d]$ are parabolas that intercept in real time (continuous blue line is for $+$ sign and dashed red line is for $-$ sign). (b) Adiabats $E_{+,-}=\pm \sqrt{[t^2 + d]^2+k^2}$ (with $k=0.5$) do not intercept in real time (continuous blue line is for $+$ sign and dashed red line is for $-$ sign) but do intercept in complex time.
 %In both the diabatic and the adiabatic limits, the probability to tunnel from the lower to the upper band -- as illustrated by the continuity of the dashed red line between an initial and a final state both in the lower band -- vanishes.
 }
\label{fig:spectrum}
\end{figure}

Performing a unitary transformation in pseudo-spin space allows one to rewrite the $2\times 2$ hamiltonian in a familiar LZ form. Let $(\sigma_x,\sigma_y,\sigma_z)\to (\sigma_z,\sigma_x,\sigma_y)$ which is realized by the unitary operator $U=\exp(i \frac{2\pi}{3}\boldsymbol{\sigma}\cdot \mathbf{n})=\frac{1}{2}(\mathbb{I}+i\sigma_x+i\sigma_y+i\sigma_z)$ where $\mathbf{n}=(1,1,1)/\sqrt{3}$. Then $H_u(t)$ becomes
\beq
H(t) =\left(\begin{array}{cc}E_1(t)&H_{12}\\H_{21}&E_2(t) \end{array} \right)=[t^2+d]\sigma_z+ k \sigma_x
\label{dimensionlessh}
\eeq
where $E_1(t)=-E_2(t)=t^2+d$ is a quadratic function of time (in contrast to the original LZ problem in which $E_1(t)=-E_2(t) \propto t$) and $H_{21}=H_{12}=k=\textrm{const}$. The quantities $d$ and $k$ are the only two relevant dimensionless parameters. The first parameter, $d$, controls the distance to the merging transition, which occurs at $d=0$. When $d<0$ there are two Dirac cones (gapless phase) and when $d>0$ there are no Dirac cones (gapped phase). The other parameter, $k$, controls how far the electron is from hitting the Dirac cones directly \cite{tk}, see Fig. \ref{fig:zenery}; it is also a measure of the adiabaticity. We call $d$ the merging gap and $k$ the perpendicular gap. The orthonormal basis $\{|1\rangle,|2\rangle\}$ in which the hamiltonian is written is called the diabatic basis. The diabatic spectrum corresponds to a negligible $k$ and is simply $E_1(t)=t^2+d$ and $E_2(t)=-E_1(t)$. It is plotted in Fig. \ref{fig:spectrum}(a) for negative $d$. It features two band crossings in real time.
%------------------------------
%------------------------------

The state of the electron at a given time $t$ is described by the bispinor $|\psi(t)\rangle$. Its time evolution is given by the Schr\"odinger equation $i\frac{d}{dt}|\psi\rangle = H(t)|\psi\rangle$.
Let us assume that initially ($t\to -\infty$), the electron is in the lower band $|\psi(-\infty)\rangle\sim |2\rangle$. Our aim is to compute the probability $P(k,d)=|\langle 1|\psi(\infty)\rangle|^2$ that it ends in the upper band ($t\to \infty$) as a function of the two parameters $k$ and $d$. As the result does not depend on the sign of $k$, we will assume that $k\geq 0$, without loss of generality.

In the following, we mathematically formulate this problem in two different bases, namely the diabatic and the adiabatic bases.

\subsection{Diabatic basis}
In the diabatic basis, we write the state at an arbitrary time as a function of two complex numbers $A_1(t)$ and $A_2(t)$:
\beq
|\psi(t)\rangle=A_1(t)e^{-i\int^t dt' E_1(t') }|1\rangle + A_2(t)e^{-i \int^t dt' E_2(t')} |2\rangle
\eeq
The time evolution is governed by the Schr\"odinger equation, which reads:
\beqn
\dot{A}_1&=&-i H_{12}A_2(t) e^{i \int^t dt' E_{12}(t')} \nonumber \\
\dot{A}_2&=&-i H_{12}^* A_1(t) e^{-i \int^t dt' E_{12}(t')}
\label{eq:diababasis}
\eeqn
where $E_{12}(t)\equiv E_1 (t) - E_2 (t)$. One therefore needs to solve this system of coupled equations with the initial conditions $A_1(-\infty)=0$  and $A_2(-\infty)=1$ (up to a global phase factor). As $|A_1(t)|^2+|A_2(t)|^2=1$ at any time, we are only interested in finding $P=|A_1(\infty)|^2$. This system of two coupled first order differential equations can also be written as a single second order differential equation for $A_1$ (or $A_2$) alone \cite{Zener}.

If the force is large, the motion of the electron is fast and $k=c_y p_y (2m_*)^{1/3}/(\hbar F)^{2/3}\ll 1$ is negligible. This is the diabatic or sudden limit. In such a limit, the electron stays in the lower state $|2\rangle$ and $P\to 0$. Indeed, when $k=0$, the tunneling probability is zero for all $d$ as a result of the conservation of the pseudo-spin $\sigma_z$, which commutes with the hamiltonian $H(t)$. This may seem surprising as it means that even when the two bands overlap the probability of interband tunneling is zero.  In particular, when $d=0$ with a single quadratic crossing point, the electron does not tunnel to the upper band even though the gap vanishes. This is actually the same phenomenon as Klein tunneling for a 1D version of the graphene bilayer, see e.g. the appendix of Ref. [\onlinecite{AllainFuchs}]. When in the gapless phase $d<0$, this may be seen as two successive perfect Klein tunnelings for a 1D massless Dirac electron: first going from the lower to the upper band with unit probability and then going down to the lower band with certainty at the second Dirac cone. When $k$ is non-zero but small, one can solve the coupled differential equations in perturbation theory as shown below, and show that the probability becomes finite.

\subsection{Adiabatic basis}
It is also useful to write the same problem in the adiabatic basis, which corresponds to diagonalizing $H(t)$ with $t$ being treated as a parameter.  We call $E_\alpha (t)=\alpha \sqrt{(t^2+d)^2+k^2}=\alpha E_+$ the adiabatic eigenenergies (plotted in Fig. \ref{fig:spectrum}(b) when $d<0$, see also Fig. \ref{fig:zenery}), where $\alpha=\pm $ is the band index, and $|\psi_\alpha(t)\rangle$ the corresponding eigenvectors. They satisfy $H(t)|\psi_\alpha(t)\rangle =E_\alpha(t)|\psi_\alpha(t)\rangle$. The angle $\theta(t)$ is defined by $\sin \theta =k/E_+$ and $\cos \theta = (t^2+d)/E_+$, which allows us to write the adiabatic eigenvectors as
\beq
|\psi_+(t)\rangle =\left(\begin{array}{c}\cos (\theta/2)\\ \sin (\theta/2) \end{array} \right); \,\, |\psi_-(t)\rangle =\left(\begin{array}{c}\sin (\theta/2)\\ -\cos (\theta/2) \end{array} \right)
\eeq
They form an orthonormal basis at each $t$. The state of the electron at any time can now be expressed in this basis as
\beq
|\psi(t)\rangle=\sum_\alpha A_\alpha (t) e^{-i\int^t dt' E_\alpha(t')}|\psi_\alpha(t)\rangle
\eeq
in terms of two unknown amplitudes $A_\alpha(t)$, which satisfy $\sum_\alpha |A_\alpha|^2=1$. The initial state is such that $A_-(-\infty)=1$ (up to a global phase factor) and $A_+(-\infty)=0$ and we are interested in $P=|A_+(\infty)|^2$. Indeed, as $t\to \pm\infty$, $\theta \approx 0$ and $|\psi_-(t)\rangle \approx -|2\rangle$ and $|\psi_+(t)\rangle \approx |1\rangle$. Therefore at both initial and final times, the adiabatic and diabatic basis coincide.

The time dependent amplitudes satisfy the following Schr\"odinger equations
\beqn
\dot{A}_+ + A_+ \langle \psi_+|\dot{\psi}_+\rangle&=&- \langle \psi_+|\dot{\psi}_- \rangle A_-  e^{i \int^t dt' E_{+-}(t')} \nonumber \\
\dot{A}_- + A_- \langle \psi_-|\dot{\psi}_-\rangle&=&- \langle \psi_-|\dot{\psi}_+ \rangle A_+  e^{-i \int^t dt' E_{+-}(t')}
\eeqn
where $E_{+-}\equiv E_+ - E_-$. As $|\dot{\psi}_{\pm}\rangle=\mp \frac{\dot{\theta}}{2}|\psi_{\mp}\rangle$, one has $\langle \psi_\pm | \dot{\psi}_\pm \rangle =0$ and $\langle \psi_- | \dot{\psi}_+ \rangle = -\langle \psi_+ | \dot{\psi}_- \rangle =-\frac{\dot{\theta}}{2}$, so that the equations become
\beqn
\dot{A}_+ &=&- \langle \psi_+|\dot{\psi}_- \rangle A_-  e^{i \int^t dt E_{+-}(t)} \nonumber \\
\dot{A}_- &=& (\langle \psi_+|\dot{\psi}_- \rangle)^* A_+  e^{-i \int^t dt E_{+-}(t)}
\label{adiabaschro}
\eeqn
and are quite similar to the ones obtained in the diabatic basis, see eq. (\ref{eq:diababasis}). They also depend on two functions of time: one is the energy difference between the two basis states $E_{+-}(t)$ (instead of $E_{12}(t)$) and the other is the coupling between these states $\langle \psi_+|\dot{\psi}_- \rangle(t)=\dot{\theta}/2$ (instead of $iH_{12}(t)$).

If the force is small, the motion of the electron is slow and $k=c_y p_y (2m_*)^{1/3}/(\hbar F)^{2/3}\gg 1$ is large. This is the adiabatic limit and $k$ can be thought of as an adiabaticity parameter. It is also the semiclassical limit as it is equivalent to $\hbar \to 0$ (in a purely classical problem, the transition probability would always be zero). In this limit, the electron stays in the state $|\psi_-(t)\rangle$, which in both limits $t\to \pm \infty$ is $\sim |2\rangle$. As a consequence the transition probability $P=|\langle 1|\psi(\infty)\rangle|^2\approx |\langle 1|\psi_-(\infty)\rangle|^2$ is also zero. When $k$ is large but finite, it is possible to compute the transition probability in perturbation theory (this time the small parameter being $1/k$) as shown below.

To summarize, both when $k\ll 1$ and $k\gg 1$, the transition probability vanishes. Away from these two limits, the probability will be non-zero. This already shows that the probability $P$ is a non monotonic function of the perpendicular gap $k$, which is in stark contrast to the LZ problem of a linear avoided band crossing. In the following, we use perturbation theory to compute the transition probability first in the diabatic and then in the adiabatic basis.

\section{St\"uckelberg theory in the gapless phase}
\begin{figure}[ht]
\begin{center}
\includegraphics[width=7cm]{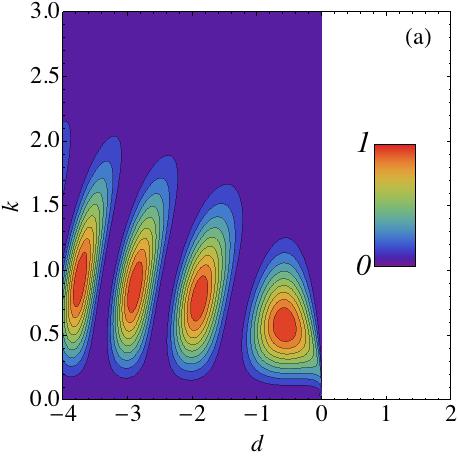}
\includegraphics[width=7cm]{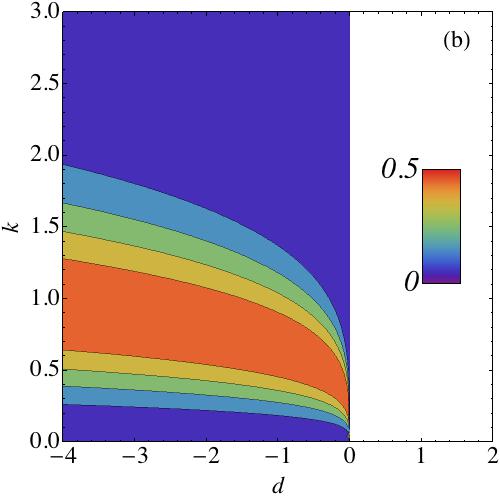}
\end{center}
 \caption{(Color online) Contour plot of the transition probability $P$  computed with the St\"uckelberg approach as a function of the merging gap $d$ and the perpendicular gap $k$.  The white region corresponds to $d\geq 0$, where the St\"uckelberg approach is not defined. (a) In the coherent case, interferences both as a function of $d$ and $k$ are clearly visible as well as the vanishing of $P$ in the $k\to 0$ and $k\to \infty$ limits. The probability varies between 0 and 1 as given by the color code (color steps corresponds to 0.1).
(b) In the incoherent case, the oscillations are washed out and the maximum probability is $1/2$ instead of $1$ in the coherent case (note the change of color scale for $P$).}
\label{fig:stueckelberg3dplot}
\end{figure}

We start by recalling the results we obtained previously using the St\"uckelberg theory in the gapless phase  ($d<0$), see the supplemental material of \cite{Lim}. We first compute the transition probability associated to the two successive LZ events, in the limit where they can be considered to be far apart (i.e. deep in the gapless phase) using the St\"uckelberg approach \cite{Stuckelberg,revueStuck,Lim}. During a single LZ event the probability amplitude to stay in the upper/lower band is $\sqrt{1-P_Z}e^{\mp i\varphi_{St}}$ where the Zener probability is $P_Z=e^{-2\pi\delta}$. The non-adiabatic phase delay $\mp \varphi_{St}$ (where $\mp$ refers to the upper/lower band) is given in terms of the Stokes phase \cite{revueStuck}
\beq
\varphi_{St}=\frac{\pi}{4} + \delta (\ln \delta -1) + \textrm{Arg}\, \Gamma(1 - i \delta)
\label{eq:stokes}
\eeq
where
\beq
\delta = \frac{k^2}{4\sqrt{|d|}}
\label{eq:delta}
\eeq
is the adiabaticity parameter in the linear band crossing problem (not to be confused with $k$, which is the adiabaticity parameter in the quadratic band crossing). In the diabatic limit, $\delta \to 0$, the Stokes phase is $\pi/4$ and it monotonically goes to zero in the adiabatic limit $\delta \to \infty$. If the sequence between the two tunneling events is coherent, the two avoided linear crossings realize a St\"uckelberg interferometer. The total probability amplitude to go from the lower to the upper band is the sum of the amplitude for two distinct paths. In the first path, the electron jumps to the upper band at the first Dirac cone and then stays in the upper band at the second, such that the amplitude is  $A_+=-\sqrt{P_Z}\times e^{i\varphi_{+}}\times \sqrt{1-P_Z}e^{-i\varphi_{St}}$ where $-\sqrt{P_Z}$ is the amplitude to jump at the first avoided band crossing and $\varphi_{+}=\int_{-\sqrt{|d|}}^{\sqrt{|d|}} dt E_+(t)$ is the phase dynamically acquired by the electron traveling in the upper band between the two Dirac cones, with $2\sqrt{|d|}$ the time needed to travel between the two Dirac points. In the second path, the electron stays in the lower band at the first Dirac cone and then jumps to the upper band at the second. The associated amplitude is   $A_-=\sqrt{1-P_Z}e^{i\varphi_{St}}\times e^{i\varphi_{-}}\times \sqrt{P_Z}$ where $\sqrt{P_Z}$ is the amplitude to jump at the second avoided band crossing and $\varphi_{-}=\int_{-\sqrt{|d|}}^{\sqrt{|d|}} dt E_-(t)$ is the dynamically acquired phase of the electron traveling in the lower band from one Dirac cone to the other. Note that the jumping amplitudes $\mp \sqrt{P_Z}$ at the two avoided crossings are opposite to each other. This is related to the fact that the linear LZ problem is not exactly the same for the two Dirac cones: indeed the local low-energy hamiltonians are slightly different just as the ones describing the two different valleys of graphene \cite{valleys}. The total probability $P =|A_+ + A_- |^2$ is therefore
%%%%%%%%%%%%%%%%%%%%%%%%%%%%%%%%%%%%%%%%%%%%%%%%%%%%%%%%%%%
 \be
P = 4 P_Z ( 1 - P_Z) \sin^2 (\frac{\varphi_{dy}}{2} + \varphi_{St}) \label{LZS}  \ee
%%%%%%%%%%%%%%%%%%%%%%%%%%%%%%%%%%%%%%%%%%%%%%%%%%%%%%%%%%%
where   $\varphi_{dy}=\varphi_{-}-\varphi_{+}$ is the dynamically accumulated phase between the two tunneling events \cite{Lim}
%%%%%%%%%%%%%%%%%%%%%%%%%%%%%%%%%%%%%%%%%%%%%%%%%%%%%%%%%%%
\be \varphi_{dy} = \int_{-\sqrt{|d|}}^{\sqrt{|d|}} E_{+-} (t) d t =4|d|^{3/2}I(\frac{k}{|d|}) \ee
%%%%%%%%%%%%%%%%%%%%%%%%%%%%%%%%%%%%%%%%%%%%%%%%%%%%%%%%%%%
written in terms of the integral $I(x)\equiv \int_0^1 du \sqrt{(u^2-1)^2+x^2}$. This probability as a function of $d$ and $k$ is plotted in Fig. \ref{fig:stueckelberg3dplot}(a).
%------------------------------
%------------------------------

If the two tunneling events are incoherent, the interferences are washed out, $\sin^2 \to 1/2$ and the probability becomes $P_\textrm{incoh}=2P_Z(1-P_Z)$ (see Fig. \ref{fig:stueckelberg3dplot}(b)).

In the limit $\delta \gg 1$, $P_Z=\exp(-2\pi\delta)\to 0$, $1-P_Z\to 1$, $\varphi_{St} \approx 0$ and $\varphi_{dy} \approx 4k \sqrt{|d|}$. Therefore
\beq
P\approx 4e^{-\frac{\pi k^2}{2\sqrt{|d|}}}\sin^2\left(2k\sqrt{|d|}\right)
\eeq
The probability goes to zero exponentially because of the large gap, as is usual for tunneling process.

In the opposite limit $\delta \ll 1$, $P_Z\to 1$, $1-P_Z\approx 2\pi \delta \to 0$, $\varphi_{St} \approx \pi/4$ and $\varphi_{dy} \approx 8|d|^{3/2}/3$. Therefore
\beq
P\approx 2\pi\frac{k^2}{\sqrt{|d|}} \sin^2\left(\frac{4}{3}|d|^{3/2}+\frac{\pi}{4} \right) \label{eq:stuckdeltasmall}
\eeq
The probability also goes to zero but as $k^2$ because of the special symmetry when $k=0$ (conservation of the pseudo-spin $\sigma_z$).

Quantitatively, the St\"uckelberg approach is valid if the Zener tunneling time $\sim \textrm{max} (\delta,\sqrt{\delta})/k$ is shorter than the time $2\sqrt{|d|}$ it takes for an electron to travel between the two Dirac points \cite{revueStuck}. This means that $-d\gg 1$ and $-d \gg k$. Therefore, one needs to be deep in the gapless phase (far from the merging) and with a perpendicular gap that is not too large.

\section{Perturbation theory in the diabatic/sudden limit}
In the diabatic basis, we perform perturbation theory in the perpendicular gap $k\ll 1$. Assuming that $A_2(t)\approx 1$ for all $t$ gives the probability $P=|A_1(\infty)|^2$ to tunnel from the lower to the upper band in terms of the amplitude
\beqn
A_1(\infty)&=&-k\int_{-\infty}^{\infty} dt A_2(t) \exp[i\int_0^t dt' E_{12}(t')]\nonumber \\
&\approx& -k\int_{-\infty}^{\infty} dt \exp[i(2 t d + 2 t^3/3)]
\label{eq:airyamp}
\eeqn
computed at first order in $k$.
The probability
\beq
P\approx 4^{2/3}\pi^2 k^2 [\textrm{Ai}(4^{1/3}d)]^2 \qquad (k\ll 1)
\label{eq:airy}
\eeq
is given in terms of the Airy function which has the following definition (when its argument $x$ is real):
\beq
\textrm{Ai}(x)\equiv \frac{1}{2\pi}\int_{-\infty}^\infty dy e^{i(\frac{1}{3}y^3+xy)}
\eeq
As it will be useful later, we also perform a saddle point analysis of the integral in eq. (\ref{eq:airyamp}) in three different limits to obtain simpler analytical results. If $d\neq 0$, there are two saddle points $t_0$ in the complex time plane. If $d>0$, $t_0=\pm i\sqrt{d}$ and only $t_0=i\sqrt{d}$ contributes, as $\textrm{Im } t_0\geq 0$ is needed. If $d<0$, $t_0=\pm \sqrt{|d|}$ and the two saddle points contribute giving rise to interferences (this is really a stationary phase approximation). If $d=0$, there is a single saddle point at $t_0=0$. The results of the saddle point approximation are
\beq
P\approx \left(\frac{2}{3}\right)^{4/3} \left(\frac{\pi}{\Gamma(2/3)}\right)^{2} k^2 \textrm{ if } |d|\ll 1
\eeq
\beq
P\approx \frac{\pi k^2}{2\sqrt{d}}e^{-\frac{8d^{3/2}}{3}}\textrm{ if } d\gg 1
\eeq
\beq
P\approx \frac{2\pi k^2}{\sqrt{|d|}}\sin^2 \left(\frac{4}{3}|d|^{3/2}+\frac{\pi}{4}\right) \textrm{ if } -d \gg 1
\eeq
The last case recovers the result of the previous section, see Eq. (\ref{eq:stuckdeltasmall}), featuring St\"uckelberg oscillations. These three limits are well known expansions of the Airy function.
%------------------------------
\begin{figure}[ht]
\begin{center}
\includegraphics[width=7cm]{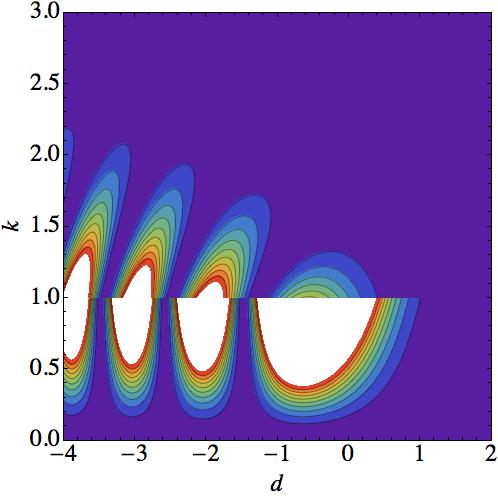}
\end{center}
 \caption{(Color online) Contour plot of the transition probability $P$  as a function of the merging gap $d$ and the perpendicular gap $k$. For small $k$, it is computed with the diabatic perturbation theory eq. (\ref{eq:airy}); while for large $k$, it is computed using adiabatic perturbation theory eq. (\ref{eq:adiabaproba}), see section V. Interferences as a function of $d$ are clearly visible in the gapless phase ($d<0$). In the gapped phase ($d>0$), the probability vanishes exponentially. White regions corresponds to the probability exceeding 1. Indeed close to $k\sim 1$, the two perturbative approaches break down. The color code is the same as in Fig. \ref{fig:stueckelberg3dplot}(a)}
\label{fig:diabapertu3dplot}
\end{figure}
%------------------------------
%------------------------------
\begin{figure}[ht]
\begin{center}
\includegraphics[width=7cm]{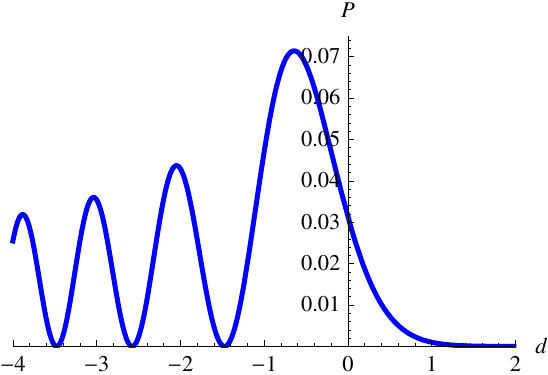}
\end{center}
 \caption{Transition probability $P$ as a function of $d$ for fixed $k=0.1$ as computed in diabatic perturbation theory, see eq. (\ref{eq:airy}). For negative argument (gapless phase, $d<0$) it shows St\"uckelberg oscillations and then decays exponentially for positive argument (gapped phase, $d>0$). There is an inflexion point right at the merging (when the argument vanishes $d=0$). }
\label{fig:Airy}
\end{figure}
%------------------------------
The transition probability $P$  of eq. (\ref{eq:airy}) is plotted in Figs. \ref{fig:diabapertu3dplot} and \ref{fig:Airy}. It goes to zero as $k \to 0$ as expected from the $\sigma_z$ conservation and increases quadratically with $k$. As a function of $d$, $P$ shows oscillations when $d<0$, which we interpret as St\"uckelberg interferences, and decreases exponentially when there is a true gap $d>0$ in the diabatic spectrum.

\section{Perturbation theory in the adiabatic/semiclassical limit}
In order to consider the opposite adiabatic limit ($k\gg 1$), we use perturbation theory in the adiabatic basis \cite{Dykhne}. From the adiabatic eigenenergies and eigenvectors, we find that $E_{+-}=2E_+$, $\langle \psi_+|\dot{\psi}_+\rangle=0$ and $\langle \psi_+|\dot{\psi}_-\rangle=\dot{\theta}/2=-k t/E_+^2$. The Schr\"odinger equations to be solved are therefore
\beqn
\dot{A}_+ &=&-\frac{\dot{\theta}}{2}A_-  e^{2i\int^t dt' E_{+}(t')} \\
\dot{A}_- &=& \frac{\dot{\theta}}{2} A_+  e^{-2i \int^t dt' E_{+}(t')}
\eeqn
with the initial conditions $A_-(-\infty)=1$ (up to a phase factor) and $A_+(-\infty)=0$. If we now assume that the coupling $\dot{\theta}$ is small, we find that $A_-(t)\approx 1$, $\dot{A}_+ \approx -\frac{\dot{\theta}}{2} e^{2i\int^t dt E_{+}(t)}$ and therefore:
\beq
A_+(\infty)\approx -\int_{-\infty}^{+\infty} dt \frac{\dot{\theta}}{2} e^{2i\int^t dt' E_{+}(t')}\equiv-\int_{-\infty}^{+\infty} dt \frac{\dot{\theta}}{2} e^{i \phi(t)}
\eeq
where $\phi(t)\equiv 2\int_{t_l}^t dt' E_{+}(t')$ is the adiabatic phase and $t_l$ is the lower bound of the phase integral, which is undecided for the moment except that it has to be a real number.

\begin{figure}[ht]
\begin{center}
\psfrag{t1}{$t_1$}
\psfrag{t4}{$t_4$}
\includegraphics[width=5cm]{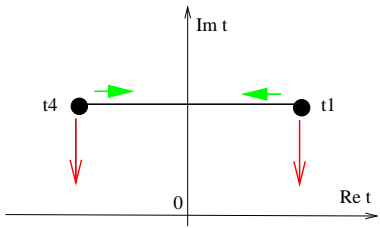}
\end{center}
 \caption{(Color online) Poles $t_1$ and $t_4=-t_1^*$ in the complex time plane. Poles are represented at the merging transition ($d=0$ such that $\textrm{arg } t_1=\pi/4=-\textrm{arg } t_4$). The short green arrows indicate their motion when $d$ increases (gapped phase $d>0$) and the long red arrows that when it decreases (gapless phase $d<0$). When the poles have a finite real part, there are oscillations in the probability; whereas a finite imaginary part implies an exponential decay of the probability. Note that there is a branch cut relating the two poles.}
\label{fig:poles}
\end{figure}

The amplitude $A_+(\infty)$ can be computed by integration in the complex plane. Firstly, there are four poles corresponding to $E_+(t)=0$ i.e. to $t^2=-d\pm ik$. Note that these are also saddle points as $\dot{\phi}(t)=2E_+$. In the following we refer to them simply as poles even when they are playing there the role of saddle points. If we write $-d+ik=\sqrt{k^2+d^2}e^{i\beta}$, which defines the angle $\beta$, the four poles are $t_1=(k^2+d^2)^{1/4} e^{i\beta/2}$, $t_2=t_1^*$, $t_3=-t_1$ and $t_4=-t_1^*$. In addition to the four poles, there are also branch cuts coming from the square root function in the exponential. The corresponding branching points are at the same position as the poles. Therefore, there is a branch cut linking $t_1$ and $t_4$ and another one linking $t_2$ to $t_3$, for example. When constructing a closed contour, one has to keep in mind this branch cut structure. Of the four poles, only $t_1$ and $t_4$ have a positive imaginary part and are therefore relevant as we want to close the integration contour in the upper plane (see Fig. \ref{fig:poles}). Since $\textrm{Im } t_1=\textrm{Im } t_4$, both poles contribute equally to the amplitude. We choose the lower bound $t_l=0$ such that $\phi(t)= 2\int_0^t dt' E_{+}(t')$. It is important to make a single choice for $t_l$ for both poles as they will interfere. As $E_+(-t)=E_+(t)$ and $t_4=-t_1^*$, we have $\phi(t_4)=-\phi(t_1)^*$. One possibility is therefore to construct a contour that encloses both these two poles and the branch cuts.
%------------------------------
%------------------------------

Another trick to perform this integral is to make a change of variable from the time $t$ to the phase $\phi$ variable, see for example  Ref. \cite{BerryMount}. The resulting integral
\beq
A_+(\infty)\approx -\frac{1}{2} \int_{-\infty}^{+\infty} d\phi \, \frac{d \theta}{d \phi} e^{i \phi}
\eeq
is over a function $(d\theta / d\phi)e^{i\phi}$ that has no branch cut anymore and only four isolated poles at $\phi_1\equiv \phi(t_1)$, $\phi(t_2)=\phi_1^*$, $\phi(t_3)=-\phi_1$ and $\phi(t_4)=-\phi_1^*$. The residue theorem can now be used with a simple contour closed in the upper complex plane of the $\phi$ variable. As the residues are $-e^{i\phi_1}/3i$ and $e^{-i\phi_1^*}/3i$, we obtain
\beq
A_+(\infty)\approx \frac{\pi}{3}(e^{i\phi_1}-e^{-i\phi_1^*})=\frac{2i\pi}{3} \sin (\textrm{Re }\phi_1) e^{-\textrm{Im } \phi_1}
\label{eq:amplitude}
\eeq
This is valid whatever the sign of $d$. Therefore
\beq
P\approx \frac{4\pi^2}{9} \sin^2 (\textrm{Re }\phi_1) e^{-2\textrm{Im } \phi_1}
\label{eq:probawrong}
\eeq
where
\beq
\phi_1\equiv \phi(t_1)=2k^{3/2}\int_0^{u_1} du \sqrt{1+(u^2+D)^2}
\label{eq:phione}
\eeq
with $D\equiv d/k$ and  $u_1\equiv t_1/\sqrt{k}=(\sqrt{\sqrt{1+D^2}-D}+i\sqrt{\sqrt{1+D^2}+D})/\sqrt{2}$.

The result we found is the first-order perturbation in the adiabatic basis. The exponential behavior is correct but not the prefactor, even in the adiabatic limit, as argued by Landau long ago \cite{LandauBook, seealsodykhne}. This is known in the literature as the ``$\pi/3$ problem'' \cite{BerryMount,Berry1990}. Actually, in the case of a single linear band crossing ($E_1=-E_2=\alpha t/2$ and $H_{12}=$ constant), which is the standard LZ problem, adiabatic perturbation theory gives $P=(\pi/3)^2 \exp(-2\pi |H_{12}|^2/\alpha)$ \cite{DavisPechukas} whereas the exact result found by Zener is $P_Z=\exp(-2\pi |H_{12}|^2/\alpha)$ \cite{Zener}. The reason for this discrepancy is well explained in \cite{Berry}: it comes from the fact that each order of the adiabatic perturbation expansion for $A_+$ contains a term of the form  $\# \exp(-\pi |H_{12}|^2/\alpha)$. Obtaining the exact factor in front of the exponential requires re-summing the whole series by keeping only the dominant -- in the adiabatic limit -- exponential behavior in each order. This series has a first term which is $\pi/3$ and a sum which is $1$ \cite{Dykhne,DavisPechukas,Berry,Berry1990}.
In the adiabatic limit, the correct pre-exponential factor in the usual LZ problem is such that
\beq
P\approx e^{-\textrm{Im} \int_{t_0^*}^{t_0}dt E_{+-}(t)}=e^{-2 \textrm{Im} \int_{t_l}^{t_0}dt E_{+-}(t)}
\eeq
which amounts to reducing the residue of the pole found in first order adiabatic perturbation theory from $\pi/3$ to $1$ in the amplitude. This can also be done in the two poles case and we find that $A_+(\infty)\approx \frac{\pi}{3}(e^{i\phi(t_1)}-e^{-i\phi(t_1)^*})$ $\to$ $e^{i\phi(t_1)}-e^{-i\phi(t_1)^*}$ so that the probability becomes, instead of eq. (\ref{eq:probawrong}),
\beq
P\approx 4 \sin^2 (\textrm{Re }\phi_1) e^{-2\textrm{Im } \phi_1} \qquad (k\gg 1)
\label{eq:adiabaproba}
\eeq
where $\phi_1(k,d)$ is given in eq. (\ref{eq:phione}). As we will later see, this result agrees very well with the exact numerical solution. It also agrees with the St\"uckelberg theory in the adiabatic limit when $P_Z\ll 1$ such that $P\approx 4P_Z \sin^2(\varphi_{dy}/2)$ where $P_Z=\exp(-\pi k^2/(2\sqrt{-d}))$ deep in the gapless phase (indeed $2e^{-2 \textrm{Im } \phi_1}\approx e^{-\pi k^2/(2\sqrt{|d|})}$). Therefore, we take this result as the correct analytical expression in the adiabatic limit.

\begin{figure}[ht]
\begin{center}
\includegraphics[width=7cm]{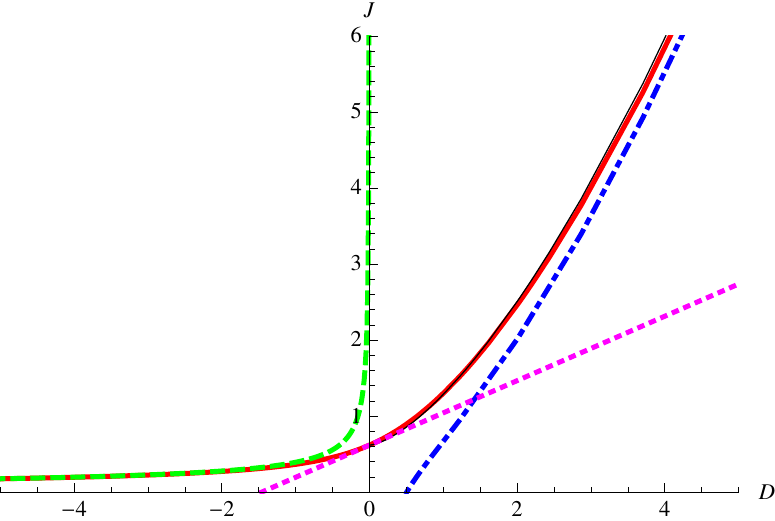}
\end{center}
 \caption{(Color online) Integral $J\equiv \textrm{Im }\int_0^{u_1} du \sqrt{1+(u^2+D)^2}$ -- giving $\textrm{Im }\phi_1=2k^{3/2}J(d/k)$ -- plotted as a function of $D$. The numerical calculation is in continuous red and is compared to different analytical results: $\pi/(8\sqrt{|D|})$ in dashed green, $\frac{\Gamma(1/4)^2}{12\sqrt{\pi}}+\frac{\pi^{3/2}}{\Gamma(1/4)^2}D$ in dotted magenta, and $2D^{3/2}/3+\ln D/(4\sqrt{D})$ in dot-dashed blue. Interpolation formula (for positive $d$) $\frac{\Gamma(1/4)^2}{12\sqrt{\pi}}+\frac{2}{3}D^{3/2}$ is in thin black.}
\label{fig:JD}
\end{figure}

We now come back to the phase $\phi_1$ given in eq. \ref{eq:phione}. As $u_1(D)=iu_1(-D)^*$, one has $\phi_1(k,d)=i\phi_1(k,-d)^*$ and therefore $\textrm{Re }\phi_1(k,d)=\textrm{Im }\phi_1(k,-d)$, which allows one to express $P$ in terms of $\textrm{Im}\phi_1$ only. The integral $J(D)\equiv \textrm{Im }\int_0^{u_1} du \sqrt{1+(u^2+D)^2}$ giving $\textrm{Im } \phi_1(k,d)=2k^{3/2}J(d/k)$ can be computed numerically for any $D$ and analytically in three limits. When $D\sim 0$, $J(D)\approx \frac{\Gamma(1/4)^2}{12\sqrt{\pi}}+\frac{\pi^{3/2}}{\Gamma(1/4)^2}D$. When $D\to \infty$, $J(D)\approx 2D^{3/2}/3+\ln D/(4\sqrt{D})$. In practise, a good approximate interpolation between $0$ and $\infty$ is $J(D)\approx \frac{\Gamma(1/4)^2}{12\sqrt{\pi}}+\frac{2}{3}D^{3/2}$. When $D\to - \infty$, $J(D)\approx \pi/(8\sqrt{|D|})$. This function $J(D)$ is plotted in figure \ref{fig:JD}.
%------------------------------
%------------------------------

From the behavior of $J(D)$, we can obtain approximate analytical results for the probability $P$ in three limits:
\beq
P\approx 4e^{-\frac{\Gamma(1/4)^2}{3\sqrt{\pi}} k^{3/2}}\sin^2(\frac{\Gamma(1/4)^2}{6\sqrt{\pi}} k^{3/2}) \textrm{ if } |d| \ll k
\eeq
\beq
P\approx 4e^{-\frac{8 d^{3/2}}{3}}\sin^2(\frac{\pi k^2}{4\sqrt{d}})  \textrm{ if } d \gg k>0
\eeq
\beq
P\approx 4e^{-\frac{\pi k^2}{2\sqrt{|d|}}}\sin^2(\frac{4 |d|^{3/2}}{3})  \textrm{ if } -d \gg k>0
\eeq
%------------------------------
\begin{figure}[ht]
\begin{center}
\includegraphics[width=7cm]{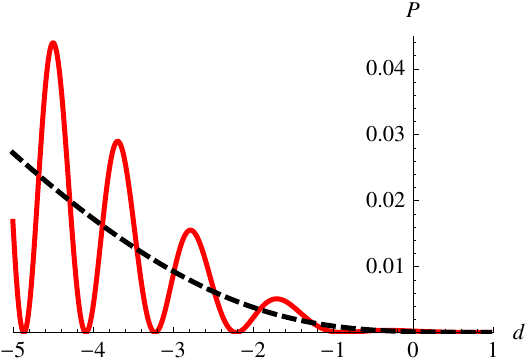}
\end{center}
 \caption{(Color online) Transition probability $P$ as a function of $d$ (at fixed $k=2.5$) computed in adiabatic perturbation theory (continuous red line), see eq. (\ref{eq:adiabaproba}). Also shown, in dashed black, is the incoherent probability $P_{incoh}$, see eq. (\ref{eq:incohadiabaproba}).}
\label{fig:AdiaProba}
\end{figure}
%------------------------------
If the contribution of the two poles add incoherently, $\sin^2 \to 1/2$, and the oscillations are washed out:
\beq
P_\textrm{incoh}\approx 2 e^{-2\textrm{Im} \phi_1}
\label{eq:incohadiabaproba}
\eeq
as in the St\"uckelberg theory when $P_Z\ll 1$.

It is interesting to discuss the motion of the two poles $t_1=(k^2+d^2)^{1/4} e^{i\beta/2}$ and $t_4=-t_1^*$, where $-d+ik=\sqrt{k^2+d^2}e^{i\beta}$, in the complex time plane as $k$ and $d$ vary \cite{tphi}. These two poles correspond to the band crossings at complex times and always exist (whatever the sign of $d$ and even at the merging). When $d=0$, $\beta/2=\pi/4$ and $\textrm{Re } t_1 = \textrm{Im } t_1$. When $d>0$, $\pi/4<\beta/2<\pi/2$ and $\textrm{Re } t_1< \textrm{Im } t_1$, the poles are close to the imaginary axis, the corresponding exponentials are essentially decaying and the probability as well. In the limit $d\to +\infty$, $\beta/2\to \pi/2$ and the two poles are on the imaginary axis. Remember that, in the diabatic limit $k\to 0$, we found a saddle point at $t_0=i\sqrt{d}$, i.e. $\beta/2 \sim \pi/2$ and $(k^2+d^2)^{1/4}\sim \sqrt{d}$. When $d<0$, $0<\beta/2<\pi/4$ and $\textrm{Re } t_1>\textrm{Im } t_1$, the poles are close to the real axis, the corresponding exponentials are essentially oscillating and the interference of the two give oscillations in the probability. In the limit $d\to -\infty$, $\beta/2\to 0$ and the two poles are on the real axis. Remember that, in the diabatic limit $k\to 0$, we found two stationary points at $t_0=\pm \sqrt{-d}$, i.e. $\beta/2 \sim 0$ and $(k^2+d^2)^{1/4}\sim \sqrt{-d}$. The approximate St\"uckelberg theory also falls in this general frame. It corresponds to a situation where $-d\gg 1,k$ such that $t_1\approx \sqrt{|d|}$ and $t_4 \approx -\sqrt{|d|}$ (there, we identified $t_1-t_4\approx 2\sqrt{|d|}$ as the time needed to travel between the two Dirac cones). We speculate that in the general case of arbitrary $k$ and $d$, there are always two separated complex poles with the same positive imaginary part. The motion of the poles in the complex $t$ plane as $d$ changes at fixed $k\neq 0$ is illustrated in Fig. \ref{fig:poles}.

\section{Modified St\"uckelberg formula}
\begin{figure}[ht]
\begin{center}
\includegraphics[width=7cm]{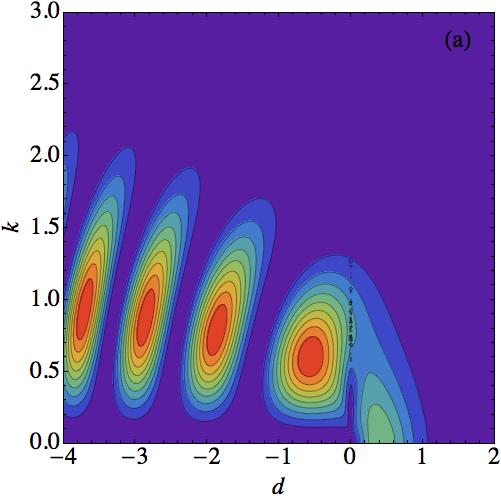}
\includegraphics[width=7cm]{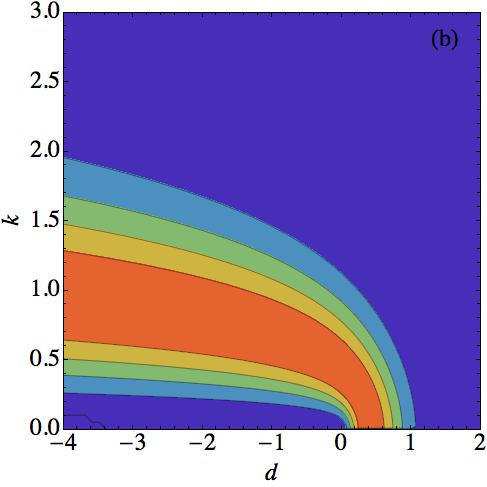}
\end{center}
 \caption{(Color online) Contour plot of the modified St\"uckelberg transition probability $P$ as a function of the merging gap $d$ and the perpendicular gap $k$.  (a) Coherent case (see eq. (\ref{eq:modstuckcont})): the probability is in between 0 and 1 (the color code is the same as in Fig. \ref{fig:stueckelberg3dplot}(a)). Note that the modified St\"uckelberg formula does not work in the ($d\geq 0,k<1$) region as the non-adiabatic phase is not properly given by the Stokes phase; (b) Incoherent case (see eq. (\ref{eq:incohmodstuck})): the probability is in between 0 and 0.5 (the color code is the same as in Fig. \ref{fig:stueckelberg3dplot}(b)).}
\label{fig:modstuckcont}
\end{figure}

In the preceding section, adiabatic perturbation theory helped us uncover a general two poles structure -- either in the complex $t$ or complex $\phi$ plane --, which leads to a total probability of the St\"uckelberg form $P=4P_S(1-P_S)\sin^2(...)$, where $P_S$ is the probability for a single avoided crossing. This should be valid for all $k$ and $d$ and not only when the spectrum is gapless. A reasonable guess (see also \cite{Suominen}) is to combine the adiabatic perturbation theory, giving the exponential weight of the two poles and their interferences in the adiabatic limit, with the St\"uckelberg approach, giving the $P_S(1-P_S)$ structure. From eqs. (\ref{LZS}) and (\ref{eq:adiabaproba}), we obtain:
\beq
P\approx 4 e^{-2\textrm{Im } \phi_1}(1-e^{-2\textrm{Im } \phi_1}) \sin^2 (\textrm{Re }\phi_1+\varphi_{na})
\label{eq:modstuckcont}
\eeq
There, $e^{i \phi_1}$ is the amplitude to tunnel for a single pole, so that $e^{-2\textrm{Im } \phi_1}$ plays the role of the Zener probability $P_Z$ for a single Dirac cone and $\textrm{Re }\phi_1$ that of $\varphi_{dy}/2$. The quantity $\varphi_{na}$ is the non-adiabatic phase acquired by a particle when it does not tunnel at a single pole -- the associated amplitude being $\sqrt{1-e^{-2\textrm{Im } \phi_1}}e^{i\varphi_{na}}$. We only know its expression in the St\"uckelberg limit ($d\ll -1, -k$), where it is given by the Stokes phase $\varphi_{St}$, see eq. (\ref{eq:stokes}) with $\delta$ as in eq. (\ref{eq:delta}). Here, we assume that $\varphi_{na}\approx \varphi_{St}$ for all $k$ and $d$, which is a reasonable approximation except when $k<1$ and $d\geq 0$. Equation (\ref{eq:modstuckcont}) should be exact both for small $k$ and negative $d$, where it recovers the St\"uckelberg result eq. (\ref{LZS}), and for large $k$, where it recovers the result of adiabatic perturbation theory eq. (\ref{eq:adiabaproba}) for all $d$. By continuity, it should also be reasonable in the intermediate region $k\sim 1$, see Fig. \ref{fig:smallk}(b). It allows one to have an approximate analytical formula that can describe the crossover from small to large $k$ at fixed $d$, see Fig. \ref{fig:negatived}. This modified St\"uckelberg probability is plotted in Fig. \ref{fig:modstuckcont}(a).  As seen, this formula is not applicable for positive $d$ and small $k$ as the relevant non-adiabatic phase is no more simply given by the Stokes phase.  In the incoherent case, the probability becomes
\beq
P_\textrm{incoh}\approx 2 e^{-2\textrm{Im } \phi_1}(1-e^{-2\textrm{Im } \phi_1})
\label{eq:incohmodstuck}
\eeq
 and is plotted in Fig. \ref{fig:modstuckcont}(b). As this incoherent probability does not depend on the partly unknown non-adiabatic phase, it should be reasonable in the whole $(d,k)$ plane.
%------------------------------
%------------------------------

\section{Numerical solution and comparison between different approaches}
\begin{figure}[ht]
\begin{center}
\includegraphics[width=7cm]{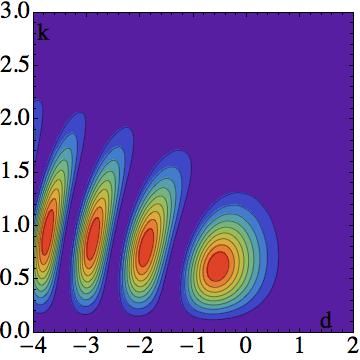}
\end{center}
 \caption{(Color online) Contour plot of the numerically computed transition probability $P$ as a function of the merging gap $d$ and the perpendicular gap $k$. Oscillations are clearly visible in the gapless phase, whereas the probability is vanishingly small in the gapped phase. The vanishing of the probability in both the diabatic $k\ll 1$ and adiabatic $k\gg 1$ limits is also visible. The color code is the same as in Fig. \ref{fig:stueckelberg3dplot}(a).}
\label{fig:num3dplot}
\end{figure}

The coupled first-order differential equations of section II, see eq.~(\ref{eq:diababasis}) and eq.~(\ref{adiabaschro}), are solved numerically. We checked that solving these equations either in the diabatic or in the adiabatic formulation gives the same answer (up to numerical errors of order $10^{-3}$ in the probability). We can therefore consider that these numerical solutions are essentially exact and use them to check the approximate analytical solutions. The probability obtained numerically for any $d$ and $k$ is shown in Fig. \ref{fig:num3dplot}. When compared to diabatic perturbation theory, the agreement is perfect for small $k\ll 1$. When compared with the St\"uckelberg theory, the agreement is very good when $d$ is very negative and $k$ not too large compared to $-d$ ($-d\gg 1$ and $-d \gg k$). It also compares very well with adiabatic perturbation theory (provided $\pi/3 \to 1$) when $k$ is large ($k\gg 1$).
%------------------------------
%------------------------------

\begin{figure}[ht]
\begin{center}
\includegraphics[width=7cm]{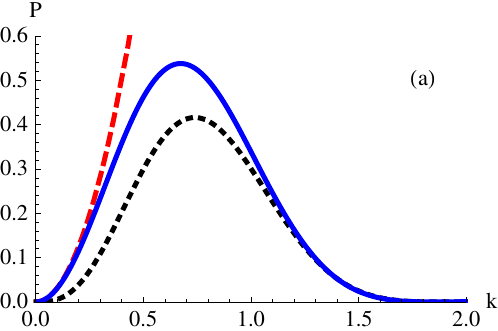}
\includegraphics[width=7cm]{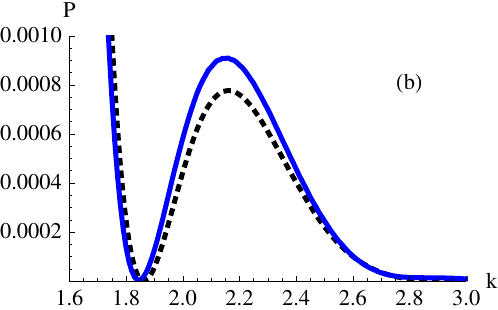}
\end{center}
 \caption{(Color online) Transition probability $P$ at the merging $d=0$ as a function of $k$. The numerically exact result is in continuous blue. The diabatic perturbative result eq. (\ref{eq:airy}) is in dashed red and the adiabatic perturbative result eq. (\ref{eq:adiabaproba}) is in dotted black. (a) $k$ between 0 and 2. (b) $k$ between 1.5 and 3 (note the change of vertical scale by a factor $10^3$): there is a tiny oscillation due to the interference between two poles.}
\label{fig:atmerging}
\end{figure}

To compare the different approaches, we first concentrate on the $d=0$ case exactly at the merging transition. The numerical solution along with the diabatic and adiabatic perturbative results are shown in Fig. \ref{fig:atmerging}. Note the excellent agreement in both the $k\to 0$ (diabatic perturbation theory) and the $k\to \infty$ limits (adiabatic perturbation theory). At large $k$, a surprising oscillation in the probability is seen both in the numerical solution and in the adiabatic perturbative result. It is surprising because the spectrum (whether diabatic $E=\pm t^2$ or adiabatic $E=\pm \sqrt{t^2+k^2}$) features at most a single real time crossing. However in complex time, the adiabatic bands cross twice. This leads to an interference between the two complex poles $t_1$ and $t_4$ and results in oscillations in the probability $P$.
%------------------------------
%------------------------------

\begin{figure}[ht]
\begin{center}
\includegraphics[width=7cm]{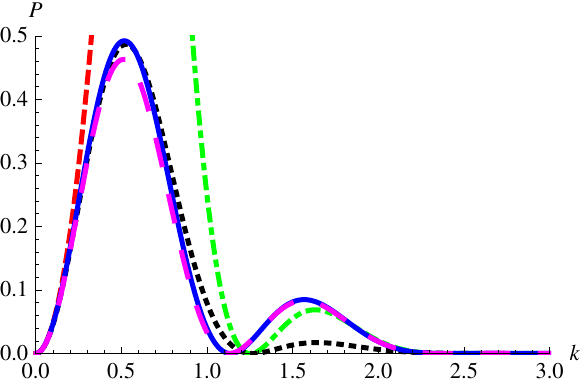}
\end{center}
 \caption{(Color online) Transition probability $P$ at fixed $d=-1$ as a function of $k$. The numerically exact result is in continuous blue, the diabatic perturbative result eq. (\ref{eq:airy}) is in dashed red, the St\"uckelberg result eq. (\ref{LZS}) is in dotted black, the adiabatic perturbative result eq. (\ref{eq:adiabaproba}) is in dot-dashed green and the modified St\"uckelberg formula eq. (\ref{eq:modstuckcont}) is in long dashed magenta.}
\label{fig:negatived}
\end{figure}

Next we consider the gapless region $d=-1$ and compare the different analytical approaches to numerics, see Fig. \ref{fig:negatived}. Note the excellent job done by the modified St\"uckelberg formula which is able to describe the whole crossover from small to large $k$. The only discrepancy with the numerical result is close to the maximum probability near $k=0.5$.
%------------------------------
%------------------------------

\begin{figure}[ht]
\begin{center}
\includegraphics[width=7cm]{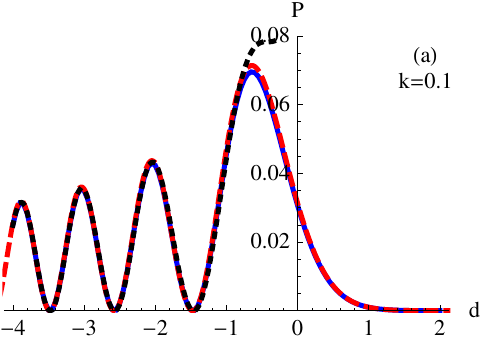}
\includegraphics[width=7cm]{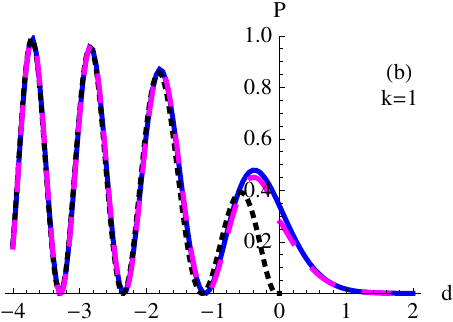}
\includegraphics[width=7cm]{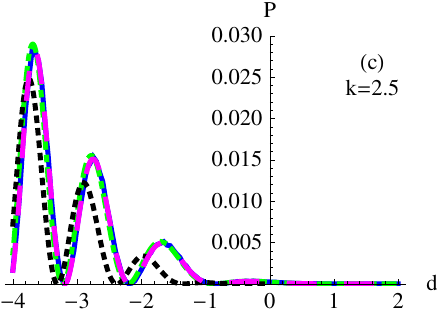}
\end{center}
 \caption{(Color online) Transition probability $P$ at fixed $k$ as a function of $d$. The numerically exact result is in continuous blue, the diabatic perturbative result eq. (\ref{eq:airy}) in dashed red, the St\"uckelberg result eq. (\ref{LZS}) in dotted black, the adiabatic perturbative result eq. (\ref{eq:adiabaproba}) in dot-dashed green and the modified St\"uckelberg formula eq. (\ref{eq:modstuckcont}) in long dashed magenta. (a) $k=0.1$ (the modified St\"uckelberg formula co\"incides with the St\"uckelberg probability when $d<0$ -- it is not shown for clarity -- and is not applicable when $d\geq 0$); (b) $k=1$; (c) $k=2.5$. Note the different probability scales in the three graphs.}
\label{fig:smallk}
\end{figure}

Then we consider small $k$ and compare numerics, the St\"uckelberg approach and diabatic perturbation theory as a function of $d$, see Fig. \ref{fig:smallk}(a). Diabatic perturbation theory agrees very well with the numerical result except for a small difference close to $d=-1$ where the probability is not small and the approximation is therefore not so good anymore. St\"uckelberg theory works very well deep in the gapless phase and its validity breaks down as one approaches the merging transition. The opposite limit of large $k$ shows that the adiabatic perturbation theory is very good (see Fig. \ref{fig:smallk}(c)). The St\"uckelberg theory works qualitatively in the gapless regime but not as well as for small $k$.
%------------------------------
%------------------------------
There are also regimes which are difficult to access analytically. This is the case for intermediate $k\sim 1$. See Fig. \ref{fig:smallk}(b) for the $k=1$ curve as a function of $d$. St\"uckelberg theory works fine but only covers the large $-d$ regime, whereas both perturbative calculations (not shown for clarity) are unreliable for intermediate $k$. The modified St\"uckelberg is qualitatively correct also when $k\sim 1$.

\section{Comparison to the experiment: absence of interferences}
Recently, an experiment with ultracold fermionic atoms in an optical lattice could study the merging transition and the inter-band tunneling of atoms performing Bloch oscillations \cite{Tarruell}. There, atoms moving in an artificial graphene-like crystal could mimic Bloch electrons in a usual solid state crystal. In Ref. \cite{Lim}, in order to understand the result of this experiment, we computed the inter-band transition probability for a single atom using the approximate St\"uckelberg theory as a function of $k$ and $d$. Then we translated these two parameters into the experimentally tunable laser intensities $V_{\bar{X}}$ and $V_X$ defined in \cite{Tarruell}. Qualitatively, $V_{\bar{X}}$ controls the merging transition and is roughly equivalent to $-d$ (called $-\Delta_*$ in \cite{Lim}), whereas $V_X$ controls the transverse gap and is equivalent to $k$ (called $c_x$ in \cite{Lim}). A last step was to average the probability over the atomic distribution of a two-dimensional degenerate Fermi gas. The agreement between theory and experiment was found to be very good: compare Fig. 4(b) in \cite{Tarruell} with Fig. 4(b) in \cite{Lim}.

However, as the St\"uckelberg theory is only valid in the gapless phase ($d<0$) and not too close to the merging transition (see Fig. \ref{fig:stueckelberg3dplot}(b)), we could only compare theory and experiment in the gapless region. Near the transition and in the gapped region, the experimental signal was vanishingly small and could not be compared with any theoretical prediction. Within the present framework, it is now possible to understand the inter-band probability very close to the merging. When looking at Fig. 4(b) of Ref. \cite{Tarruell} in detail, one sees that the red line of maximum transition probability -- which lies essentially in the gapless region -- actually crosses the merging line and slightly extends in the gapped region at very small $V_X$. This line qualitatively corresponds to $P_Z =1/2$ such that $P=1/2$. Such a behavior is found in our calculations as well: see Fig. \ref{fig:modstuckcont}(b), where the orange region of maximum probability (between 0.4 and 0.5) lies essentially in the gapless ($d<0$) region but also slightly extends to the gapped region ($d>0$) reaching $d\sim 0.4$ when $k\to 0$.

We now consider the merging point ($d=0$) and study both the inter-band probability $P_2$ for the motion in the direction where two Dirac cones are hit ($x$ direction) and that, $P_1$, in the perpendicular direction in which a single cone is hit ($y$ direction) \cite{xy}. In the present paper, we concentrate on $P_2$ -- which is called $P(d,k)$ -- as $P_1$ is simply given by the usual LZ formula and was studied in detail in Ref. \cite{Lim}. The merging point is special in the sense that the spectrum is gapless and features a single contact point, which is linearly dispersing in the $p_y$ direction and quadratically in the $p_x$ direction \cite{Dietl}. The LZ formula gives $P_1=\exp[-\pi (p_x^2/(2m_*))^2/(\hbar c_y F)]$. In the coherent case, the numerical solution of section VII gives $P_2=P(d=0,k)$ where $k= c_y p_y (2m_*)^{1/3}/(\hbar F)^{2/3}$ as plotted in Fig. \ref{fig:atmerging} (see the continuous blue line) and, in the incoherent case, the modified St\"uckelberg eq. (\ref{eq:incohmodstuck}) gives $P_2\approx 2\exp(-\frac{\Gamma(1/4)^2}{3\sqrt{\pi}}k^{3/2})[1-\exp(-\frac{\Gamma(1/4)^2}{3\sqrt{\pi}}k^{3/2})]$. The probability $P_1$ depends on $p_x$ and varies between 0 and 1, whereas $P_2$ depends on $p_y$ (i.e. on $k$) and varies between 0 and $\sim 0.55$ (coherent case, see Fig. \ref{fig:atmerging}(a)) or $0.5$ (incoherent case). The ratio $P_2/P_1$ can therefore take any positive value depending on what are the relevant $p_x$ and $p_y$ values. The latter depend on the size of the atomic cloud and on the way the averaging over the atomic cloud is done. For example, for a single atom $p_x=p_y=0$ giving $P_1=1$ and $P_2=0$ so that $P_2/P_1=0$. In particular, there is no reason for this ratio to take the simple value 0.5 \cite{Tarruell}. We have performed averaging over various atomic cloud sizes comparable to that in the ETH Z\"urich experiment and find that $\langle P_2 \rangle / \langle P_1 \rangle$ can vary between 0 and $\sim 0.7$.

One very striking experimental fact remains to be explained: the agreement is actually obtained with the {\it incoherent} inter-band probability (see e.g. Fig. \ref{fig:modstuckcont}(b)) rather than with the coherent probability (see e.g. Fig. \ref{fig:num3dplot}). In other words, St\"uckelberg oscillations (interferences) are not observed in the experiment, whereas they are predicted. Here we would like to discuss this specific point in more details. The absence of interferences could be due to (i) decoherence, (ii) blurring or (iii) washing out because of some averaging process. (i) Decoherence is unlikely in a cold atom experiment with almost {\it non-interacting} fermions. We estimate the decoherence time due to spontaneous emission following Ref. \cite{Kolovsky}. It is roughly given by $1/\tilde{\gamma} = (\delta/\Omega)^2/\gamma\sim 10^3$ s where $\gamma \sim 6$ MHz is the natural line width for the relevant transition of $^{40}$K, $\delta \sim 108$ THz is the detuning and $\Omega \sim 1$ GHz is the Rabi frequency estimated from $\hbar \Omega^2/\delta \sim E_R$, where $E_R\sim 4.4$ kHz is the recoil energy. It is much longer than the experimental time, therefore ruling out decoherence as a possible mechanism to explain the absence of interferences. (ii) Blurring of the interferences could also occur because of the detection process using a finite pixel size. We checked that possibility and found that the pixel size is small enough that it should allow experimentalists to resolve the interferences. (iii) We are left with the possibility of washing out of the oscillations due to several averaging processes. We included averaging over a two-dimensional atomic distribution in reciprocal space, which only resulted in slightly smoothing the oscillations (compare Fig. \ref{fig:num3dplot} here and Fig. 4(d) in \cite{Lim}). However, the atomic cloud in the experiment was actually not two but three-dimensional, even though the optical lattice was two-dimensional. The atomic gas was indeed confined by an anisotropic three dimensional harmonic trap but very far from the regime where one of the direction of motion would be frozen. This means that the system is best seen as a bunch of parallel one-dimensional tubes, each tube corresponding to a single site of a two-dimensional honeycomb-like lattice. The atoms hop in a kind of tight-binding lattice in the $xy$ plane (except for a weak harmonic trap $m\omega_x^2 x^2/2 + m \omega_y^2 y^2/2$) and are almost free to move in the $z$ direction (except for a weak harmonic trap $m\omega_z^2 z^2/2$). The period of the harmonic motion in the $z$ direction $2\pi/\omega_z$ is very long compared to the time an atom spends in the St\"uckelberg interferometer $\sim 2\sqrt{|d|} t_{car}$, where $t_{car}=(2m^* \hbar)^{1/3}/F^{2/3} $. One can therefore think that an atom moves in the interferometer at an almost constant $z$. However, because of the finite laser waist, the laser intensities are inhomogeneous, so that atoms at different $z$ experience a slightly different optical lattice. In other words, the parameters $d$ and $k$ of the universal hamiltonian are slightly $z$-dependent. As seen in Fig. \ref{fig:num3dplot} for example, the interferences in the inter-band probability $P$ are essentially a function of $d$ (and not so much of $k$), with a fringe spacing of roughly $\delta d \sim 1$  (which is the same as saying that $\delta \Delta_*\sim 0.04 E_R$ \cite{Lim}). From the experimental conditions of Ref. \cite{Tarruell}, we estimate a laser waist of $\sim 150$ microns and a cloud radius of $\sim 30$ microns in the $z$ direction (half of the tube's length) so that the parameter $d$ varies by roughly $0.7$ between the center and the edge of the atomic cloud. As this is comparable to the spacing between a dark and a bright fringe, it should be enough to wash out the oscillations. In the experiment, the inter-band probability is automatically averaged over the third spatial direction, i.e. along the tubes axis. Therefore, we think that the averaging over the third spatial direction could be responsible for the absence of the oscillations in the inter-band probability. An alternative explanation for the disappearance of the oscillations was very recently proposed in Ref. \cite{Uehlinger}. It is based on the spatial inhomogeneity of the applied force in the 2D plane, which also leads to averaging and washing out of the probability fringes.

By breaking the inversion symmetry of the lattice, it is also possible to induce a mass to the Dirac fermions, i.e. to gap the Dirac cones when $d<0$ \cite{Tarruell}. Such a situation is easily incorporated in our theory by a simple mapping $k\to \sqrt{k^2+g^2}$. The hamilonian (\ref{dimensionlessh}) becomes $H=[t^2+d]\sigma_z+k\sigma_x+g\sigma_y$ where $g$ is the (dimensionless) mass gap induced by inversion symmetry breaking. The inter-band transition probability $\mathcal{P}(d,k,g)$ when $g\neq 0$ is simply related to that $P(d,k)$ at $g=0$  by $\mathcal{P}(d,k,g)=\mathcal{P}(d,\sqrt{k^2+g^2},0)=P(d,\sqrt{k^2+g^2})$. This mapping is easily found by looking at the coupled differential equations (\ref{eq:diababasis}), in which $H_{12}=k$ becomes $H_{12}=k-ig=\sqrt{k^2+g^2}e^{i\gamma}$ where $\gamma \equiv \textrm{Arg }(k-ig)$. The phase $\gamma$ is time independent and can be gauged away so that only the modulus of $k-ig$ matters and $H_{12}$ becomes $\sqrt{k^2+g^2}$.

\section{Conclusion}
Inspired by a recent experiment probing the merging transition of Dirac cones via Bloch-Zener oscillations of ultracold fermionic atoms \cite{Tarruell,Lim}, we have studied inter-band tunneling for a quadratic band crossing. The latter problem depends on two dimensionless parameters, which are the merging gap $d$ and the perpendicular gap $k$. We computed the probability $P$ for a particle to tunnel from the lower to the upper band as a function of $k$ and $d$. Qualitatively, the probability oscillates as a function of $d$ in the gapless phase and decays exponentially in the gapped phase. The oscillations are a result of St\"uckelberg interferences. As a function of $k$, the probability shows quite an unusual non-monotonic behavior: $P$ vanishes exponentially in the adiabatic/semiclassical limit (large $k$), which is the expected tunneling behavior in the large gap limit, but it vanishes also in the opposite diabatic/sudden limit (small $k$) as a result of a special symmetry. Indeed, when $k=0$, the conservation of the pseudo-spin $\sigma_z$ implies that $P$ vanishes. When $k\neq 0$, this symmetry is broken and, quite counter-intuitively, opening of a gap leads first to a quadratic increase of the probability to tunnel between the bands. In addition, when $k\gg 1$, there are oscillations of $P$ (as a function of both $k$ and $d$) whatever the sign of $d$. These are due to interferences between two poles in the complex time plane. The latter exist not only in the presence of Dirac points (gapless phase) but also in the gapped phase (in which case the bands do cross but at times with a finite imaginary part).

The probability $P$ of inter-band tunneling was calculated using different methods. To summarize: the numerically exact solution of the time-dependent Schr\"odinger equation is given in section VI. We also used approximate analytical techniques to compute $P$: for small $k\ll 1$ and arbitrary $d$, we used diabatic perturbation theory, see eq. (\ref{eq:airy}). For negative $d\ll -1$ and small $k\ll -d$, we employed the St\"uckelberg approach, see eq. (\ref{LZS}). And for large $k\gg 1$, we used adiabatic perturbation theory, see eq. (\ref{eq:adiabaproba}). For intermediate $k$'s, we have no exact analytical prediction but an approximate modified St\"uckelberg formula, see eq. (\ref{eq:modstuckcont}), that compares well to the numerics in the whole negative $d$ region and also for large $k$ and positive $d$ (adiabatic regime). Using the tools we have developed, it should be possible to compute the inter-band tunneling probability for many two-bands hamiltonians.

{\it Note added}: After completion of the present work, we became aware of related articles in the context of atomic collisions, in which a parabolic level crossing problem was studied, see Ref. \cite{Suominen}. The specific case $d=0$ (exactly at the merging transition) was also very recently analyzed in \cite{LehtoSuominen}, where it is called parabolic level glancing.

\begin{acknowledgments}
We thank Fr\'ed\'eric Jean Marcel Pi\'echon for many useful discussions. We acknowledge support from the Nanosim Graphene project under grant No. ANR-09-NANO-016-01.
\end{acknowledgments}


\begin{thebibliography}{99}
\bibitem{xy}Compared to Refs. \cite{Tarruell,Lim}, the $x$ and $y$ axis have been exchanged in order to match the notations of \cite{MontambauxUH}.
\bibitem{nosinglecone}We are not interested here in the motion in the perpendicular direction -- when the electron hits a single Dirac cone -- as this corresponds exactly to the original LZ problem of a linear band crossing.
\bibitem{tk}Note that $t$ is essentially $p_x$, $k$ is essentially $p_y$ and the Dirac cones are located at $(p_x=\pm \sqrt{|d|},p_y=0)$.
\bibitem{valleys}In the notations of Wittig \cite{Wittig}, in which $H(t)=(\alpha t/2) \sigma_z + H_{12}\sigma_x$, going from one valley to the other amounts to $\sigma_z\to -\sigma_z$.
\bibitem{seealsodykhne}See also Dykhne \cite{Dykhne} from which we quote: ``It is easy to show that [it] yields the correct order of magnitude of the result (correct order of the exponential function), but an incorrect factor preceding the exponential function.''.
\bibitem{tphi}Equivalently, one could discuss the motion of the $\phi_1$ and $\phi_4=-\phi_1^*$ poles in the complex phase plane.
\end{thebibliography}
\end{document}